\documentclass[a4paper]{article}

\usepackage{icrc2013}

\title{Muon Identification with VERITAS using the Hough Transform}

\shorttitle{Muon Identification with VERITAS using the Hough Transform}

\authors{
Jonathan Tyler$^{1}$,
for the VERITAS Collaboration.
}

\afiliations{
$^1$ McGill University, Department of Physics \\
}

\email{jonathan.tyler@physics.mcgill.ca}

\abstract{Imaging atmospheric Cherenkov telescope (IACT) arrays such
  as VERITAS are used for ground-based very high-energy gamma-ray
  astronomy. This is accomplished by the detection and analysis of the
  Cherenkov light produced by gamma-ray-initiated atmospheric air
  showers. IACTs also detect the Cherenkov light emitted by individual
  muons. Identification of these muons is useful because their
  Cherenkov light can be used to calibrate the telescopes. Muons
  create characteristic annular patterns in the cameras of IACTs,
  which may be identified using parametrization algorithms. One such
  algorithm, the Hough transform, has been successfully used to
  identify muons in VERITAS data. Details of this technique are
  presented here, including results regarding its effectiveness.}


\begin{document}
\maketitle

\section{VERITAS}

\begin{figure}[t]
\centering
\includegraphics[width=0.49\textwidth]{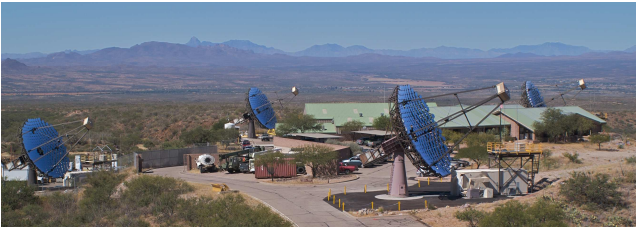}
\caption{The VERITAS array.}
\label{veritas_array}
\end{figure}

VERITAS (shown in figure~\ref{veritas_array}) is an array of four 12m
imaging atmospheric Cherenkov telescopes (IACTs) located at the base
of Mount Hopkins in southern Arizona. Very high energy gamma rays
interact with nuclei in the atmosphere, producing extensive air
showers. The particles in the air showers move faster than the speed
of light in air, emitting Cherenkov light. This light is collected and
focused with large segmented reflectors onto  arrays of light
sensitive photomultiplier tubes (PMTs) called cameras. The PMTs
produce signals that are proportional to the amount of light
detected. These signals are converted into digital information,
providing images of the air showers. The directions of the incident
gamma rays are determined by the geometry of the images. The energies
of the gamma rays are related to the geometry of the images and the
amount of light detected~\cite{bib:valcarcel}. In order to measure the
energies of the incident gamma rays, the intensity of the signals
recorded must be related to the amount of light
detected. This relationship is a poorly constrained parameter in the
energy calibration of IACTs. Muons can be used as calibrated light
sources to better determine this relationship~\cite{bib:vacanti}.

\section{Muons}

\begin{figure}[t]
\centering
\includegraphics[width=0.48\textwidth]{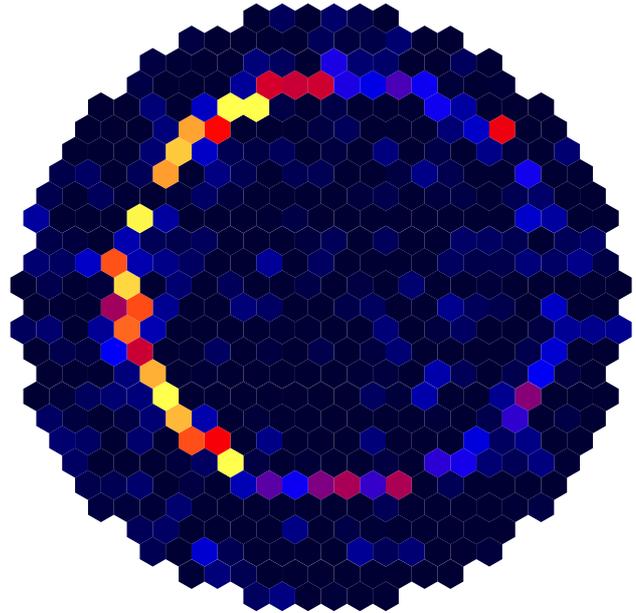}
\caption{A muon image with $b<R$ seen by VERITAS.\newline Figure courtesy of S. Fegan \& V. Vassiliev.}
\label{muonfigure}
\end{figure}

\begin{figure}[t]
\centering
\includegraphics[width=0.49\textwidth]{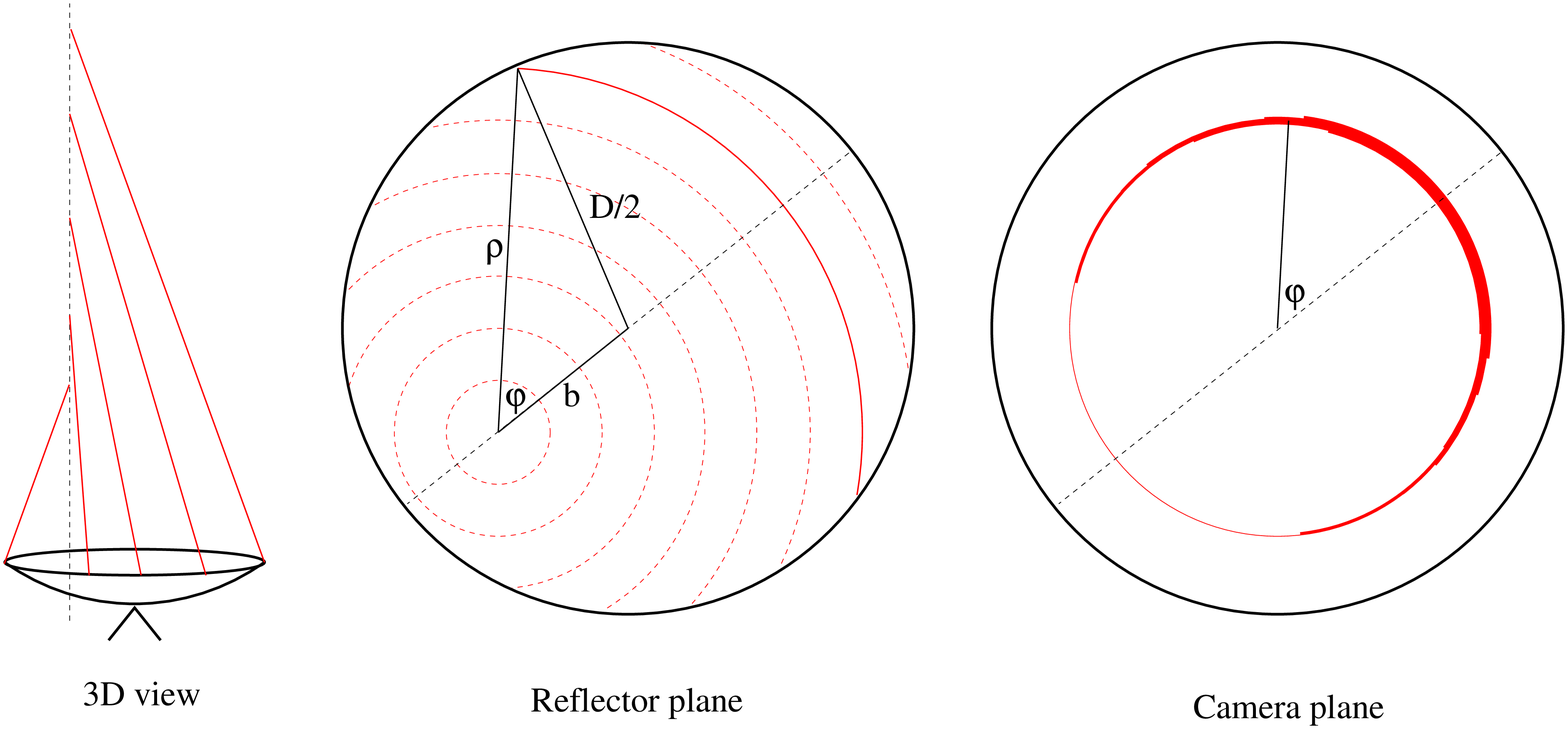}
\caption{Geometry of an on-axis muon with $b<R$.\newline Figure courtesy of S. Fegan \& V. Vassiliev.}
\label{muongeometry}
\end{figure}

Muons emit Cherenkov light in a cone at a nearly constant angle as
they propagate through the lower atmosphere. This creates annular patterns in
IACT cameras, with radii determined by the Cherenkov angles of
the muons~\cite{bib:vacanti}. Muons propagating parallel to the
optical axis (on-axis muons) that hit the center of the reflector
(impact parameter $b=0$) appear as complete rings with azimuthally
symmetric light distributions in the cameras. On-axis muons with $b \neq 0$
but less than the radius of the reflector $R$ appear as complete rings
with azimuthally dependent light distributions in the cameras as shown
in figures~\ref{muonfigure} and~\ref{muongeometry}. Distant on-axis
muons with $b>R$ appear as incomplete rings (arcs) in the
cameras. Muons propagating at angles not parallel to the optical axis
appear as rings or arcs with centers offset from the center of the
cameras. The amount of light produced by muons with known Cherenkov
angles and impact parameters is well understood. Therefore muons can
be used to calibrate the energy measurements of the
telescopes. Since muons produce characteristic
patterns, these images can be parametrized using feature finding
algorithms such as the Hough transform.

\section{Hough transform}

The Hough transform is an algorithm used for parametrizing assumed
shapes (in this case, circles) in digital images. This algorithm
converts the problem of iteratively fitting a circle in image space
into a problem of finding the best parameters for a circle in a
parameter space. Each parameter must have a limited range and
resolution defined by the user. Therefore, each pixel in a digital
image is intersected by a finite number of circular
parametrizations. Each of the circles describes a point in a 3D
parameter space $P(x,y,r)$, where $(x,y)$ is the center of the circle
and $r$ is the radius~\cite{bib:tsui}. The Hough transform employs a
histogram called an accumulator array to accumulate votes in the
parameter space. This is accomplished by adding the intensity values
of each pixel in the image to the bins of the accumulator array that
correspond to the circles that pass through those pixels. Once the
accumulator array has been filled, the best circular parametrization
of the image corresponds to the coordinates of the bin with the
highest number of votes~\cite{bib:tsui}. A lookup table was
constructed consisting of lists of circular parametrizations
associated with each pixel of the VERITAS cameras. This was
accomplished by determining which pixels were intersected by various
circular parametrizations. The locations of the centers of each PMT in
the VERITAS cameras were used as the location of the centers of the
circles, and the radii of the circles consisted of values from 3 PMT
diameters to 11 PMT diameters, incremented by a third of a PMT
diameter. This choice of parametrizations resulted in 12475
distinct circles being used to generate the lookup table. This lookup
table was used to perform the Hough transform on VERITAS events.

\section{Muon identification parameters}

\begin{figure}[t]
\centering
\resizebox{0.50\textwidth}{!}{\includegraphics[trim=0cm 10.1cm 0cm 0cm, clip=true]{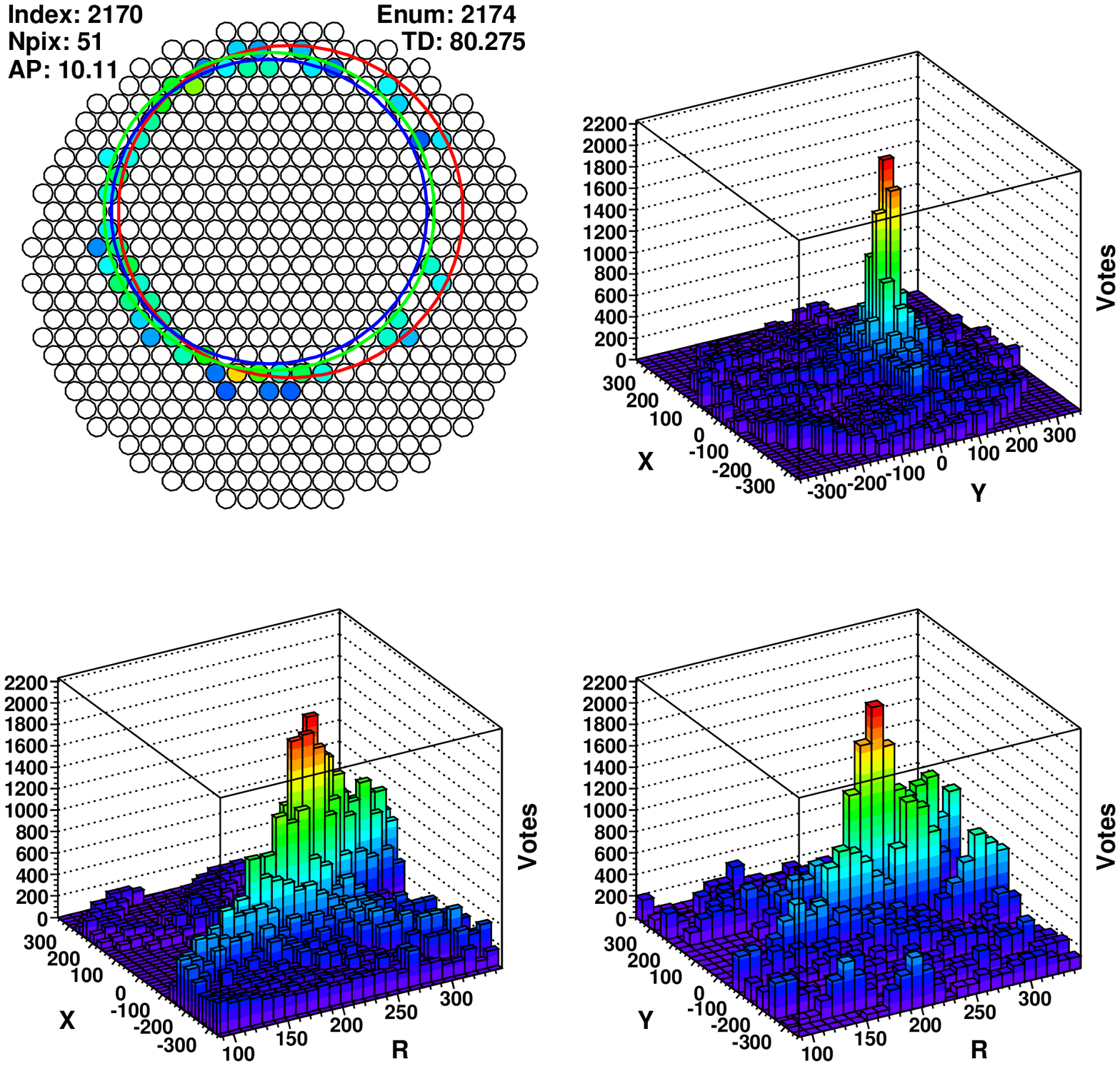}}
\resizebox{0.50\textwidth}{!}{\includegraphics[trim=0cm 10.1cm 0cm 0cm, clip=true]{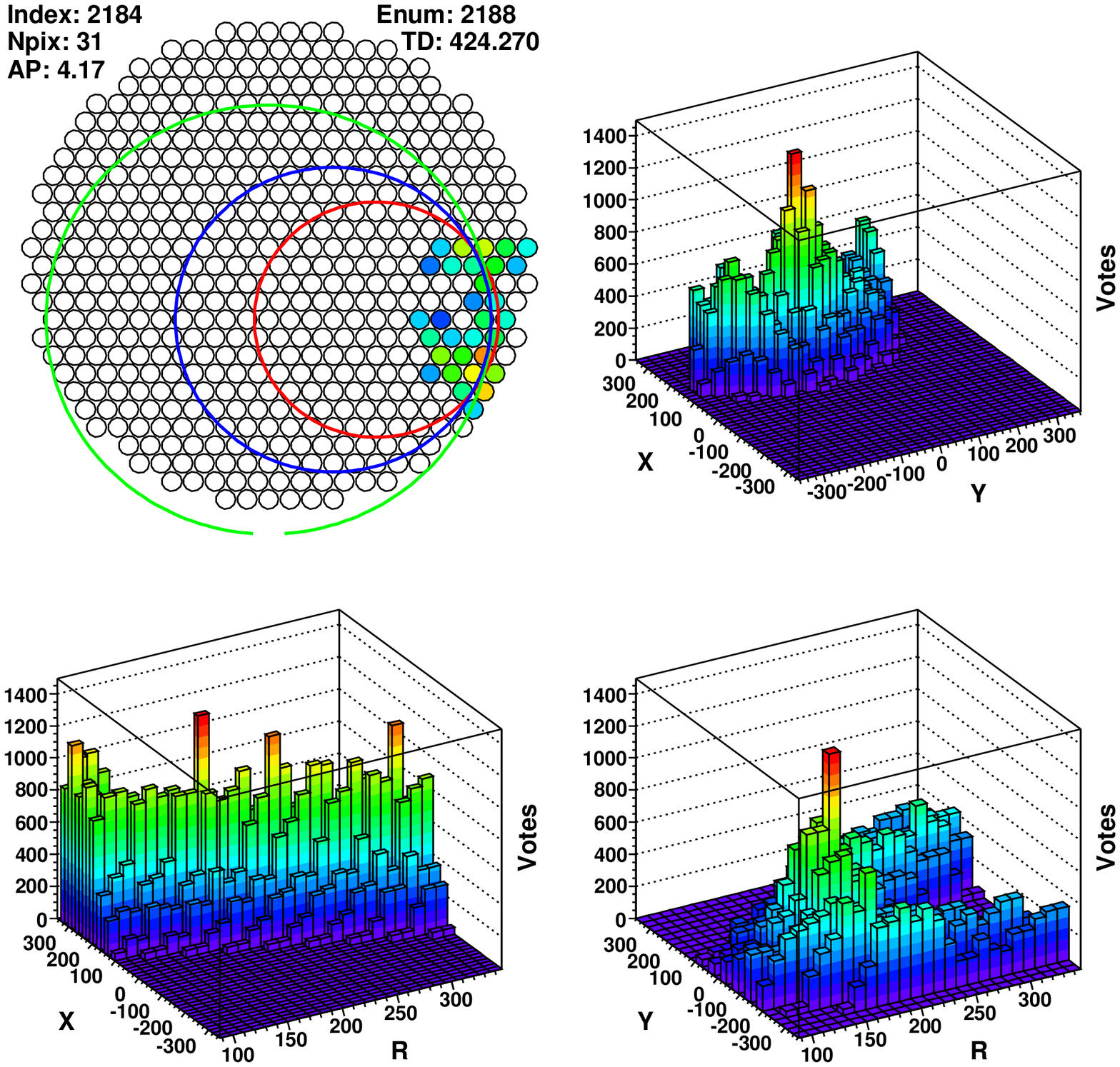}}
\caption{Pixel patterns and accumulator array projections for two events seen by VERITAS.
Top: a muon event. Bottom: a non-muon event.}
\label{events}
\end{figure}

Figure~\ref{events} shows the pixel intensity patterns (left) and 2D parameter
space projections in the $(x,y)$ plane (right) for a muon event (top) and a
non-muon event (bottom). The red, green and blue circles superimposed
over the pixel patterns correspond to the coordinates of the bins of
the accumulator array with the highest, second highest and third
highest values. For the muon event, the three best parametrizations
trace the pixel pattern quite well and a sharp peak can be seen in
parameter space projection. For the non-muon event, the three best
parametrizations differ significantly in center locations and radii
and a less peaked distribution can be seen in the parameter space
projection. These features were used to motivate the first two muon identification
parameters described below:\\

\noindent The $AP$ parameter: the value of the bin of the accumulator
array with the most votes divided by the average non-zero bin
value. Specifically:\\ 

\begin{center}
$AP = \frac{\displaystyle Largest \ bin\ value\ }{\left(
    \frac{\displaystyle Sum\ of\ all\ bin\ values}{\displaystyle
      Number\ of\ non-zero\ bins}\ \right)}$\\
\hbox{}
\end{center}

\noindent The $AP$ parameter can be thought of as a measure of the strength
or signal to noise ratio of the best parametrization of the
event. Since muon events produce sharp peaks in the accumulator array,
they should have large $AP$ values.\\

\noindent The $TD$ parameter: the sum of the distances in the parameter space
between the three best parametrizations of the event. Specifically, if
$(x_1,y_1,r_1),\ (x_2,y_2,r_2)$ and $(x_3,y_3,r_3)$ represent the
best, second best and third best parametrizations of the event,
then:\\ 

\begin{center}
$TD = D_{12} + D_{13} + D_{23}$\\*
\hbox{}
where,\\*
\hbox{}
$D_{12}=\sqrt{(x_1-x_2)^2+(y_1-y_2)^2+(r_1-r_2)^2}$\\ 
$D_{13}=\sqrt{(x_1-x_3)^2+(y_1-y_3)^2+(r_1-r_3)^2}$\\ 
$D_{23}=\sqrt{(x_2-x_3)^2+(y_2-y_3)^2+(r_2-r_3)^2}$\\
\hbox{}
\end{center}

\noindent The $TD$ parameter can be thought of as a measure of the unanimity of
the parametrizations, or the continuity of the parameter space
distribution. Since the three best parametrizations are similar for
muon events, these events should have small $TD$ values.\\

\noindent The $Npix$ parameter: the number of pixels with non-zero values after standard image processing is applied.\\

\section{Cuts and results}

In order to test the effectiveness of the parameters described in the
previous section for muon identification, 22774 events from the same
run were visually inspected and categorized. These events were labeled
muons, non-muons or ambiguous. Events with fewer than 10 hit pixels
were not categorized due to the fact that circular patterns were
difficult to identify in those images. 1516 events were categorized as muons, 17027 events were
categorized as non-muons and 4231 events were categorized as
ambiguous. The cuts on the muon identification parameters were
optimized on the first half of the visually categorized events so that
no non-muon events passed. The cuts that resulted in the greatest
number of muons were found to be:

\begin{center}
$AP\ >\ 0.011\times TD\ +\ 6.6$\\ 
$TD\ <\ 182$\\ 
$10 \leq Npix \leq 79$
\end{center}

\begin{figure}[t]
\centering
\resizebox{0.23833\textwidth}{!}{\includegraphics{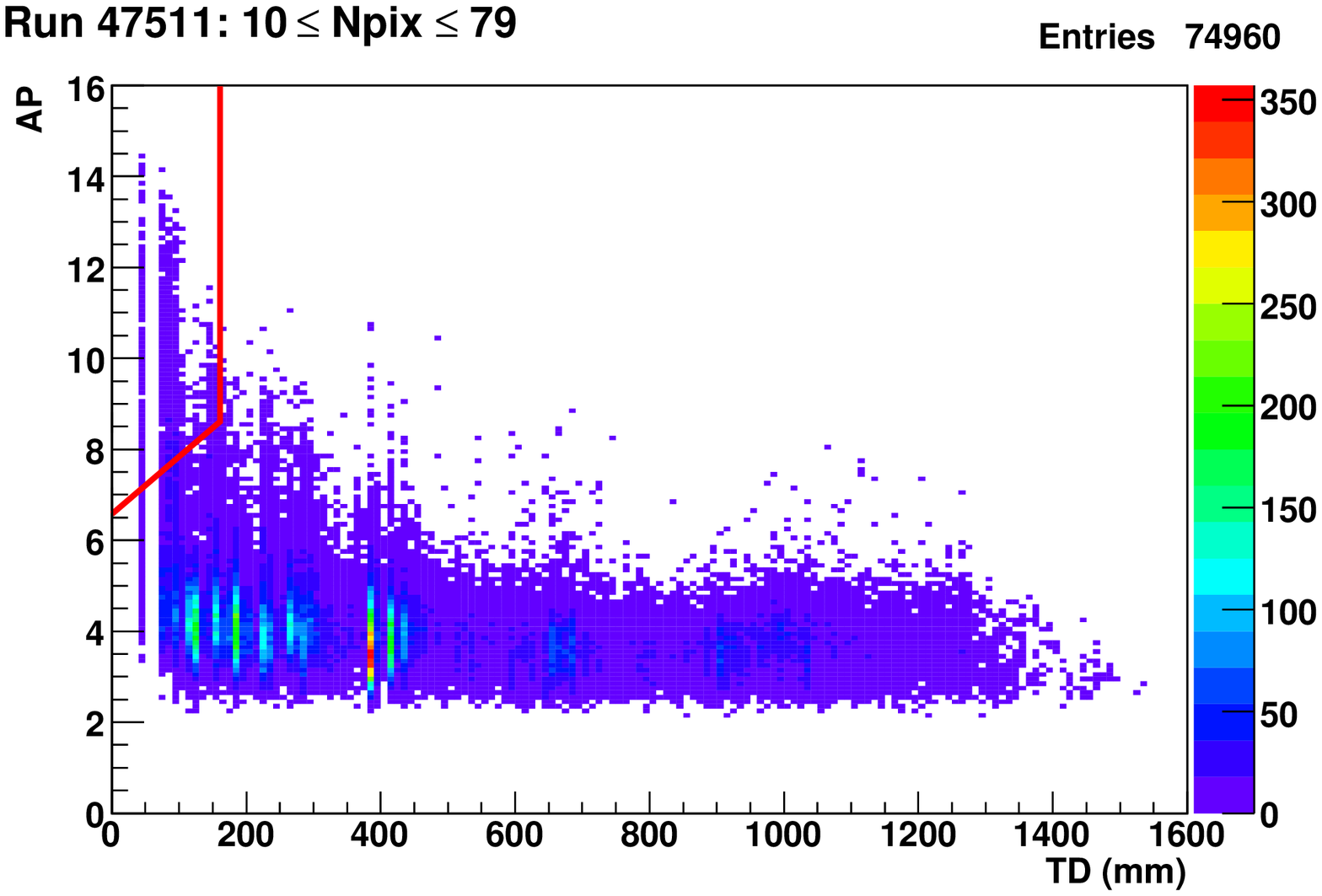}}
\resizebox{0.23833\textwidth}{!}{\includegraphics{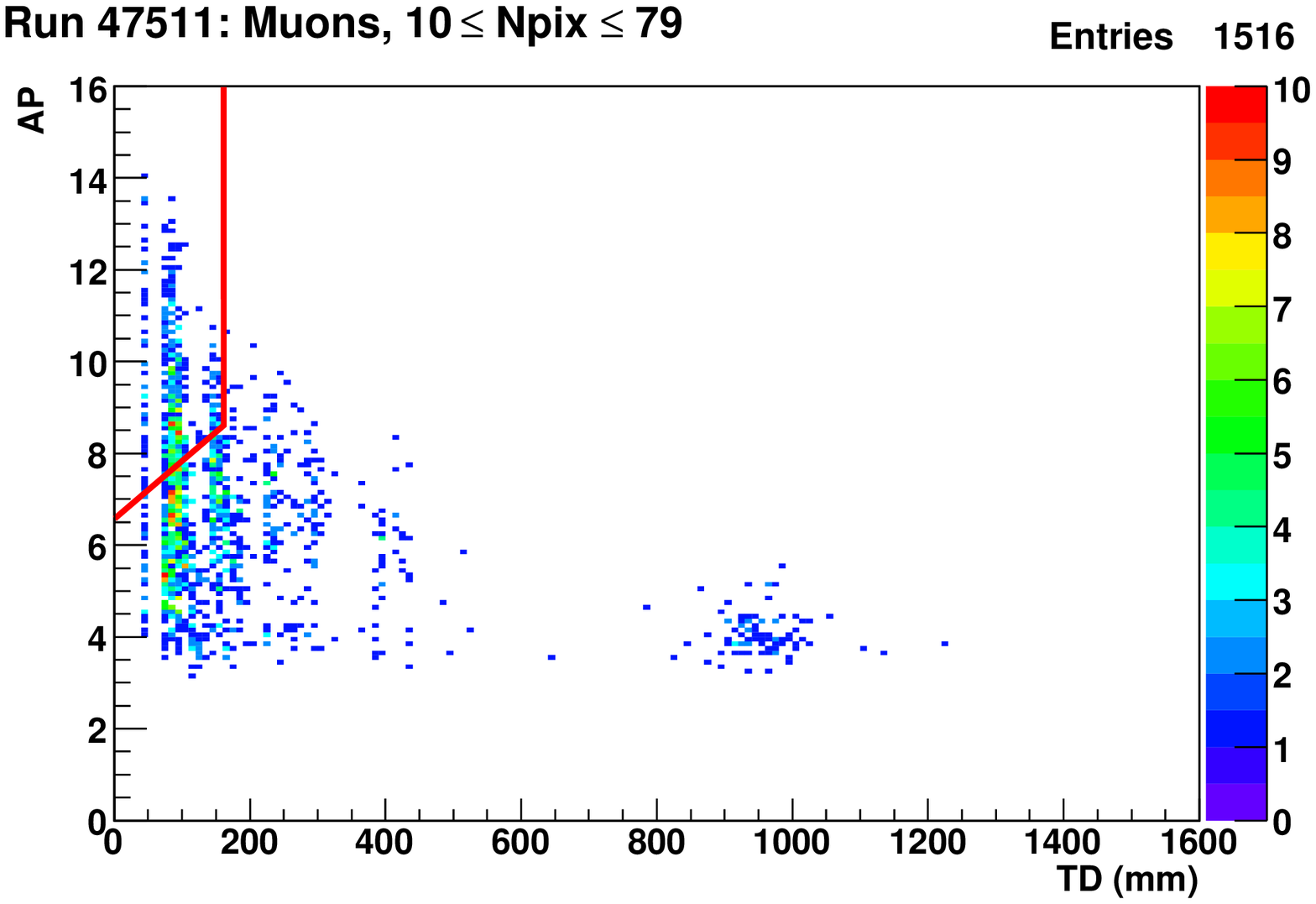}}
\resizebox{0.23833\textwidth}{!}{\includegraphics{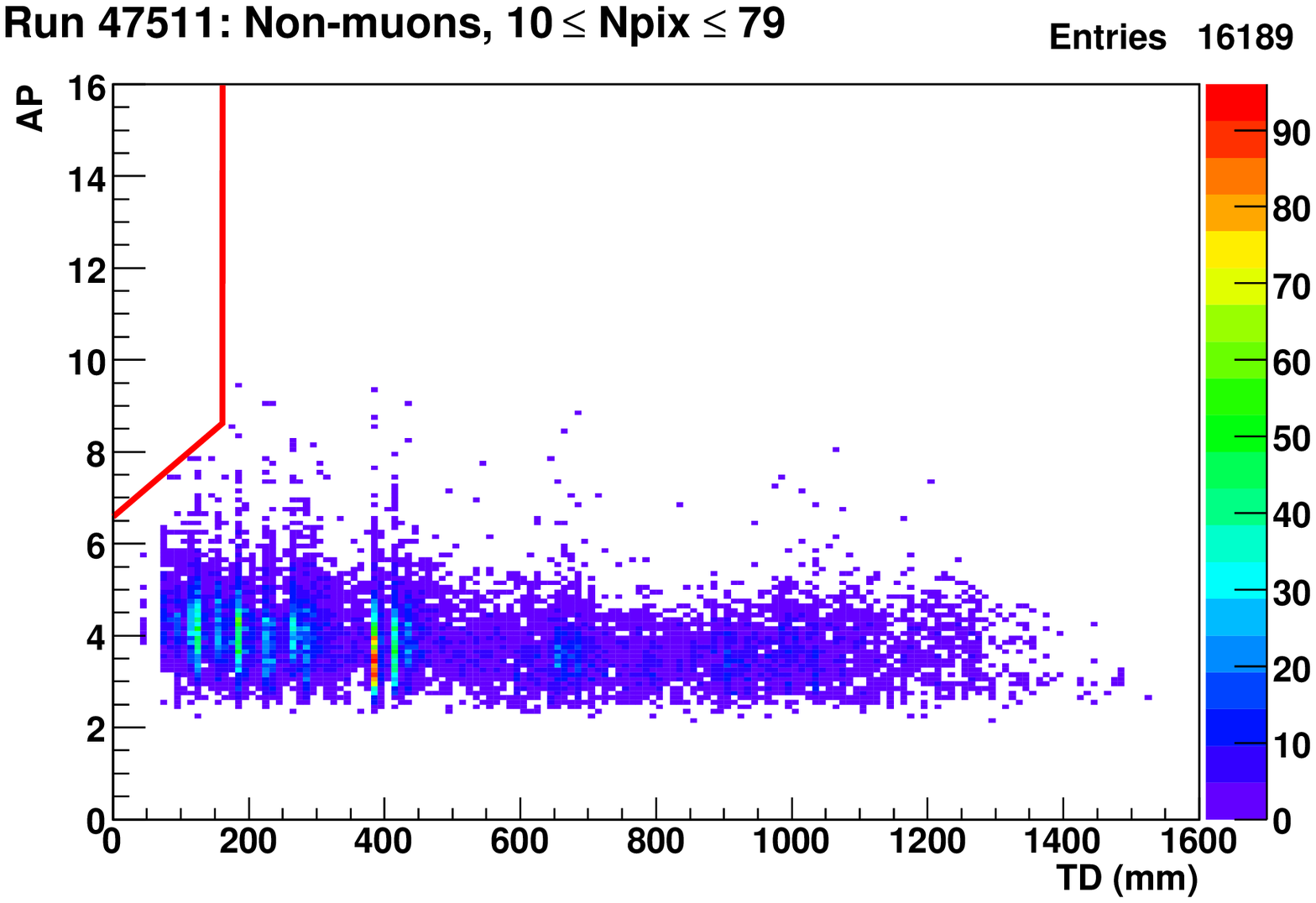}}
\resizebox{0.23833\textwidth}{!}{\includegraphics{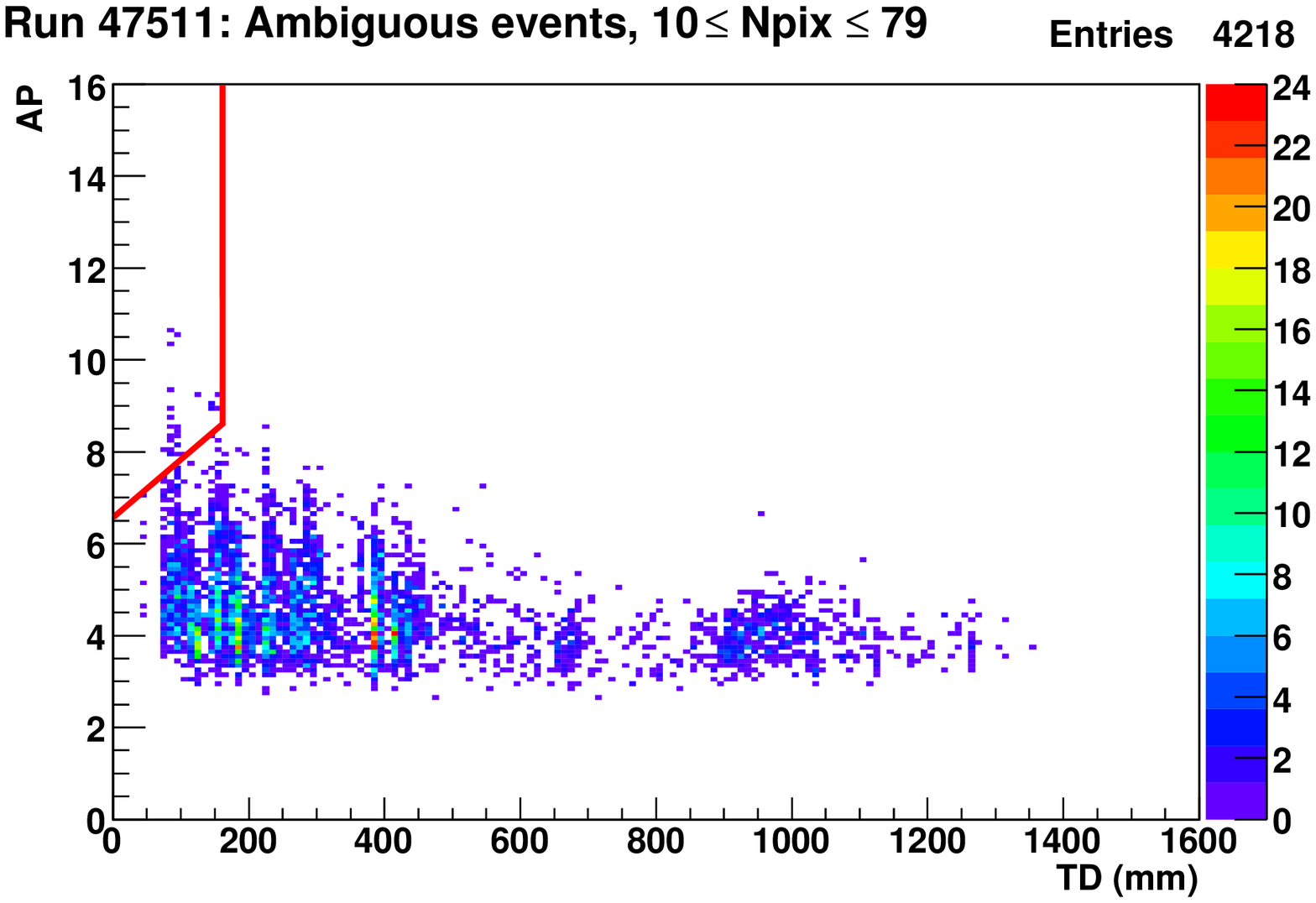}}
\caption{The $AP/TD$ distribution for $10 \leq Npix \leq 79$ in run
  47511 for different categories of events. The red lines indicate the cuts on $AP$ and $TD$.}
\label{parameterspace}
\end{figure}

Figure~\ref{parameterspace} shows the $AP/TD$ distribution for each
category of event with $10 \leq Npix \leq 79$. The upper-left plot
shows all events, the upper-right plot shows muon events, the
lower-left plot shows non-muon events and the lower-right plot shows
ambiguous events. The red lines indicate the cuts on $AP$ and
$TD$. The results of applying these cuts to the first half of the
visually categorized events are shown in
table~\ref{table:cutsrangeresults}. These cuts were then applied to
the other half of the visually categorized events and were found to
produce a pure muon sample with an estimated efficiency of
approximately 29 percent. The results of applying the cuts to the
second half of the visually categorized events are shown in
table~\ref{table:otherrangeresults}. These cuts were found to produce
highly pure muon samples when applied to other runs, as shown in
table~\ref{table:otherruns}.

\begin{table}
\begin{center}
\begin{tabular}{ | l | l | l | l | l |}
\hline & Total & Muons & Non-muons & Ambiguous \\ 
\hline Before & 10921 & 795 & 7918 & 2208 \\ 
\hline After & 239 & 225 & 0 & 14\\ 
\hline
\end{tabular}
\end{center}
\caption{The distribution of the first half of the visually
  categorized events before and after muon identification cuts.}\label{table:cutsrangeresults}
\end{table}

\begin{table}
\begin{center}
\begin{tabular}{ | l | l | l | l | l |}
\hline & Total & Muons & Non-muons & Ambiguous \\ 
\hline Before & 11853 & 721 & 9109 & 2023 \\ 
\hline After & 228 & 210 & 0 & 18
\\ \hline
\end{tabular}
\end{center}
\caption{The distribution of the second half of the visually
  categorized events before and after muon identification cuts.}\label{table:otherrangeresults}
\end{table}

\begin{table}
\begin{center}
\begin{tabular}{ | c | l | l | l |}
\hline Run & Events & Passing cuts & False positives\\ 
\hline 47511 & 274991 & 1617 & 5 \\ 
\hline 40839 & 184048 & 1105 & 2 \\ 
\hline 40840 & 166224 & 730 & 4 \\ 
\hline 40841 & 184451 & 1101 & 3 \\ 
\hline
\end{tabular}
\end{center}
\caption{The number of events scanned, number of events passing cuts and number of false positives (non-muons identified by eye but passing cuts) for four runs.}\label{table:otherruns}
\end{table}

\section{Conclusions}

The Hough transform was found to be effective at parametrizing the
circular pixel patterns produced by muons in the VERITAS cameras. The
cuts on the muon identification parameters obtained from the
accumulator array were optimized using the visually categorized events
and found to produce highly pure muon samples when applied to other
runs. This technique is currently being implemented in the VERITAS
offline analysis software. Upon completion, the
technique will be used to identify muons for calibration work. Future
research will involve improving the technique by investigating the
efficiencies of different muon identification parameters as well as
assessing the usefulness of the Hough transform algorithm for event
reconstruction.

\vspace*{0.5cm}
\footnotesize{{\bf Acknowledgments:}{ The author gratefully
    acknowledges the help of Ken Ragan, David Hanna, Andrew McCann,
    Micheal McCutcheon, Gernot Maier, Roxanne Guenette, Sean Griffin,
    Gordana Tesic, Simon Archambault, David Staszak, Jean-Francois
    Rajotte and Paul Mercure. This research is supported by grants
    from the U.S. Department of Energy Office of Science, the
    U.S. National Science Foundation and the Smithsonian Institution,
    by NSERC in Canada, by Science Foundation Ireland (SFI
    10/RFP/AST2748) and by STFC in the U.K. We acknowledge the
    excellent work of the technical support staff at the Fred Lawrence
    Whipple Observatory and at the collaborating institutions in the
    construction and operation of the instrument.}}


\end{document}